\documentclass[journal]{IEEEtran}
\usepackage{amssymb,stmaryrd,amsmath,amsfonts,rotating}
\usepackage[noadjust]{cite} \usepackage{color} 
\usepackage[vflt]{floatflt} \usepackage{epic}
\usepackage{color}
\newcommand{\bb}{\underline{b}}
\newcommand{\HH}{\underline{H}}
\newcommand{\hh}{\underline{{\sf h}}}
\newcommand{\RR}{\underline{R}}
\newcommand{\ww}{\underline{w}} 
\newcommand{\x}{\underline{x}}
\newcommand{\y}{\underline{y}}

\RequirePackage{bbm} 
\definecolor{TODO}{rgb}{0.6,0.6,0.6} 

\definecolor{TOCHECK}{rgb}{0.8,0.8,0.8} 

\newtheorem{theorem}{Theorem}
\newcommand{\btheo}{\begin{theorem}}
\newcommand{\etheo}{\end{theorem}}
\newcommand{\bproof}{\begin{proof}}
\newcommand{\eproof}{\end{proof}}
\newtheorem{definition}[theorem]{Definition}
\newcommand{\bdefi}{\begin{definition}}
\newcommand{\edefi}{\end{definition}}
\newtheorem{fact}[theorem]{Fact}
\newcommand{\bprop}{\begin{fact}}
\newcommand{\eprop}{\end{fact}}
\newtheorem{corollary}[theorem]{Corollary}
\newcommand{\bcor}{\begin{corollary}}
\newcommand{\ecor}{\end{corollary}}
\newtheorem{example}[theorem]{Example}
\newcommand{\bex}{\begin{example}}
\newcommand{\eex}{\end{example}}
\newtheorem{lemma}[theorem]{Lemma}
\newcommand{\blemma}{\begin{lemma}}
\newcommand{\elemma}{\end{lemma}}
\newtheorem{remark}[theorem]{Remark}
\newcommand{\bremark}{\begin{remark}}
\newcommand{\eremark}{\end{remark}}
\newtheorem{conj}[theorem]{Conjecture}
\newcommand{\bconj}{\begin{conj}}
\newcommand{\econj}{\end{conj}}

\def\0{{\tt 0}} 
\def\1{{\tt 1}} 
\def\?{{\tt *}} 
\newcommand{\qed}{{\hfill \footnotesize $\blacksquare$}}

 \allowdisplaybreaks
\begin{document} 
\title{Linear Programming based Detectors for Two-Dimensional
Intersymbol Interference Channels} 
\author{\authorblockN{Shrinivas Kudekar\authorrefmark{1}\authorrefmark{2},  Jason K. Johnson\authorrefmark{1} and Misha Chertkov\authorrefmark{1}\authorrefmark{2}\\ } \authorblockA{\authorrefmark{1}
Center for Nonlinear Studies \\ \& Theoretical Division T-4 \\
Los Alamos National Laboratory, Los Alamos NM, USA. \\
Email: \{skudekar, jasonj, chertkov\}@lanl.gov}\\
\authorblockA{\authorrefmark{2} New Mexico Consortium, Los Alamos, NM, USA.}
}

\maketitle
\begin{abstract}
We present and study linear programming based detectors for two-dimensional intersymbol
interference channels. Interesting instances of two-dimensional intersymbol
interference channels are magnetic storage, optical storage and Wyner's cellular
network model. 

We show that the optimal maximum a posteriori detection in such channels lends
itself to a natural linear programming based sub-optimal detector. We call this
 the Pairwise linear program detector. Our experiments show that the
Pairwise linear program detector performs poorly. We then propose two  
methods to strengthen our detector. These detectors are based on systematically
enhancing the Pairwise linear program. The first one, the Block linear program detector adds higher
order potential functions in an {\em exhaustive} manner, as constraints, to  the Pairwise linear program
detector.  We show by experiments that the Block linear program detector has
performance close to the optimal detector. We then develop another detector by
 {\em adaptively} adding frustrated cycles to the Pairwise linear program detector.
Empirically, this detector also has performance close to the optimal one and
turns out to be less complex then the Block linear program detector.  
\end{abstract}

\section{Introduction}

In this paper we consider detection of binary data in the presence of
two-dimensional intersymbol interference (2D-ISI).  Many important systems like
magnetic and optical storage are modeled as 2D-ISI channel models.  With an
increasing demand for larger storage in smaller sizes, the traditional
one-dimensional storage devices fall short. Thus there is a need for considering 2D storage devices.

2D systems will naturally suffer from 2D ISI. One such 2D storage system is the
TwoDOS (two-dimensional optical storage) \cite{Co03, ImmCo03}.  Detection in
TwoDOS reduces to detection on a 2D lattice or grid.  A detailed survey of the
2D ISI detection (and coding techniques) is given in \cite{Sie06, Kur08}.

It is known that the Vitterbi decoder achieves maximum likelihood sequence
detection \cite{For72} for detection in 1D ISI. For finite memory channels one
can thus achieve optimal detection in 1D ISI in linear time.   In general, it
is known that the 2D ISI detection problem (with additive Gaussian noise) is
NP-complete \cite{OR06}. As a consequence, there has been a lot of work in
reducing the complexity of detectors. There has been a lot of work on
developing low-complexity trellis-based detectors \cite{MaWo03a, MaWo03b,
ChAnCh01,Kris98,Weeks00,HekCoImm07}. In \cite{WuSuSiIn03, SiSuInWu02, SiSu05,
ShShSha05, ShWeShWe04} belief propagation (BP) based detectors are used for the
2D ISI channel. It was observed that the loopy BP detector performed poorly due
to the presence of many small loops. Using a joint detection and coding (turbo
equalization), loopy BP provided noise thresholds \cite{SiSu05}. In
\cite{ShWeShWe04} a generalized belief-propagation (GBP) channel detector is
shown, experimentally, to have near-optimal bit-error-rate by considering regions of size
$3\times3$. 

\subsection{Our Contributions} In this work we propose linear programming (LP)
based channel detectors.  As was observed in papers mentioned before, the
detection problem can be formulated as an inference problem on graphical
models.  We first formulate the natural LP based on the pairwise potentials of
the factor graph.    We show by experiments that (similar to loopy BP detector)
this LP performs poorly. We then propose two methods to improve the detector
based on enhancing the LP. The first one, the Block LP detector adds higher
order potential functions in an exhaustive manner, as constraints, to  the
Pairwise LP detector.    The second detector identifies frustrated cycles (see
Section~\ref{sec:LPwithFC}), when the Pairwise LP produces a fractional solution,
and adaptively adds them to the LP, which then enables us to recover the
correct information word. We show empirically that the new detectors have a
block-error performance close to the optimal one. Furthermore, the second
detector turns out to be less complex than the Block LP detector.

\section{Channel Model and Optimal Detection} 

\subsection{Channel Model: Uncoded Transmission}
We begin by describing the channel model. Consider an $N\times N$ grid. 
Let each point, $(i,j)$ $1\leq i,j \leq N$, on the grid represent an
 information bit taking value in $\{+1, -1\}$. We consider uncoded transmission. Thus the
information word belongs is $\{+1,-1\}^{N^2}$.  
 We denote by $\x$ the transmitted word and $\y$ as the received sequence. Both have length equal to $N^2$. 
 The information bit is first observed through a 2-dimensional linear filter and
then additive white Gaussian noise ($\mathcal{N}(0,\sigma^2)$) is added to get the final noisy observation
of the bit. More precisely, the 2D ISI channel model, we consider, is given by,
\begin{align}\label{eq:ISIchannel}
y_{k,l} = x_{k,l} + w_{k,l} + \sum_{(i,j)\in \partial(k,l)} h_{i,j}x_{i,j},
\end{align}
where ISI interaction strength is given by $h_{i,j}$ (strength less than 1) and $\partial (k,l)$ denotes the neighborhood of $(k,l)$.
Also notice that the central bit, $x_{k,l}$, has coefficient equal to 1 so 
that the bit under detection has the dominant contribution.
 
\subsection{Interference pattern}
 We consider the nearest-neighbor interaction, specified by a 4-neighborhood\footnote{Although it seems
the hexagonal interaction is the design choice for the TwoDOS system because of
its higher density \cite{Co03} we perform our experiments on the 
4-neighborhood for demonstrating our methods.}. 
 Furthermore, we will consider a periodic boundary.    Let us illustrate
 the ISI interactions  with an example.  
\begin{example}\label{ex:grid5interaction}  Figure~\ref{fig:ISIpatterns_graphicalmodel} shows a $5\times 5$ square grid
 with circles denoting the bits.  The bit at $(3,2)$ interacts
with four of its neighbors at $(2,2), (4,2), (3,3), (3,1)$.
Also shown is the periodic nature of our interactions. The bit on the boundary,
$(5,4)$, interacts with $(5,5), (5,3), (4,4), (1,4)$. Similar periodic
interactions are also present (but not shown in the figure) for information
bits which belong to the top-most row (they have one interaction with a bt on
the bottom-most row). We consider periodic grid interaction so that we can rule
out any boundary effects which would influence the LP detector. 
\begin{figure}[htp] \centering
\input{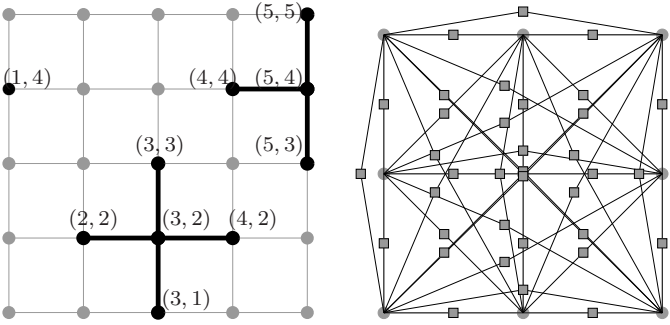}
\vspace{-0.5cm}
\caption{\label{fig:ISIpatterns_graphicalmodel} The figure on the left shows the bit at $(3,2)$  interacting with its 4
neighbors, $(2,2), (4,2), (3,3), (3,1)$. Also shown is the interaction of a bit in the boundary.  The figure on the right shows the factor graph for the
2D ISI detection problem with 9  bits and  36
potential functions denoted by squares. Each potential function is
a pairwise interaction with strength between the nodes $i$ and $j$  given by $R_{ij}x_i x_j$.
Although we do not shown them, to each
 there is a singleton potential function associated to each bit.
} 
\end{figure}
\end{example}
In our experiments we will consider uniform ISI coefficients, i.e., $h_{i,j}$ is same for all $i, j$ to illustrate our
methods.  We further assume that the channel is
perfectly known at the receiver.
In vector form the ISI channel can be written as 
$\y = \HH\,\x + \ww,$
where $\HH$ is the ISI matrix of all $h_{i,j}$. Different $\HH$ can model different applications. If the ISI coefficients are taken from a Gaussian distribution, then we can model Wyner's cellular network \cite{ShWeShWe04}. 

\subsection{Optimal Detection and the Integer Program}\label{sec:IP}
We denote by $p(\y \vert \x)$ the transition pdf of the channel. 
Let us consider the optimal or maximum a posteriori (MAP) detection on the 2D ISI channel.  
We have 
\begin{align*}
\underline{\hat{x}} & = \text{argmax}_{\x\in\{\pm 1\}^{N^2}}
\,\,\text{exp}\Big(-\frac1{2\sigma^2}(\sum_{i>j} R_{ij}x_ix_j - \sum_i {\sf h}_i
x_i)\Big),
\end{align*}
where $\RR = \HH^{T}\HH$, $\hh = \HH^T\y$ (see \cite{ShWeShWe04} for details).

Therefore the optimal detection problem reduces to solving the Integer program (IP), 
\begin{align}\label{eq:IP}
&\min_{\x \in \{\pm 1\}^{N^2}} \sum_{i>j} R_{ij} x_i x_j - \sum_{i} {\sf h}_i x_i.
\end{align}

\begin{remark}
The matrix $\RR$ introduces next-to-neighbor interactions. Hence the above model
is not planar. Above we have replaced the notation $\{x_{i,j}\}_{(i,j)\in
[1,N]\times[1,N]}$ by $\{x_i\}_{i\in [1,N^2]}$.
The figure on the right in Figure~\ref{fig:ISIpatterns_graphicalmodel} shows the factor graph of \eqref{eq:IP}. 
\end{remark}

\section{Main Results: LP Based Detectors}  An advantage of LP detectors over
GBP detectors is that the LP provides a MAP certificate. More precisely, if the
LP outputs an integer solution, then it must also be a solution to the IP and
hence LP does MAP decoding in this case.  However, in general, the IP is
NP-hard and the output of the LP can be fractional. This implies that there is
a gap in the LP approximation.  In this situation the LP relaxation provides a
lower bound (if we are considering the minimization of the objective function)
to the value of the IP.

We now provide an LP based on the pairwise potential
functions and call this the Pairwise LP. This is analogous to applying loopy
BP.   

\subsection{Pairwise LP}
We can relax the above IP to
\begin{align}
&\min_{\bb} \sum_{i>j} \sum_{x_i, x_j} R_{ij} x_i x_j b_{ij}(x_i, x_j) - \sum_i\sum_{x_i}
{\sf h}_i x_i b_i(x_i) \nonumber \\
& \text{s.t.} \quad \forall i>j:\quad \sum_{x_i, x_j}b_{ij}(x_i, x_j)=1, \label{cond1} \nonumber \\
& \forall i>j \quad \forall x_i, x_j:\quad
b_i(x_i)=\sum_{x_j}b_{ij}(x_i, x_j) \nonumber \\ 
& \hspace{3.0cm} b_j(x_j)=\sum_{x_i}b_{ij}(x_i,
x_j). \nonumber \\
& 0\leq b_i(x_i) \leq 1, \,\, \forall i, \quad\quad 0\leq b_{ij}(x_i, x_j) \leq 1, \,\, \forall i,j. 
\nonumber
\end{align}
Here $b_i(x_i)$ and $b_{ij}(x_i, x_j)$ represent the beliefs of $x_i$ and
$x_ix_j$ respectively. See \cite{FWK05} on how to derive the LP in terms of beliefs. 

\subsection{Experiments with Pairwise LP}\label{sec:exptwithPLP} 
Throughout the paper we will consider only low interference regime, i.e., $h_{i,j}$ are low. 
 Consider a $9\times 9$
grid and $h_{i,j}=0.2$ uniformly for all $1\leq i,j\leq N$. We allow the noise,
$\sigma$, to vary from $0.1$ to $1.0$ at an interval of $0.1$. We
run 2000 trials for each value of $\sigma$. In each
trial an information word $\x$ is picked u.a.r from $\{\pm 1\}^{81}$ and is combined
with a random noise configuration, $\ww$, to generate the observations $\y$. Then $\y$
and $\HH$ are fed to the Pairwise LP. If the output equals the transmitted
information word, then we declare success, else there is an error. We plot  
the word-error-rate (WER) versus SNR$=10\log_{10}((4\cdot0.2^2+1)/\sigma^2)$. 

\begin{figure}[htp] \centering
\input{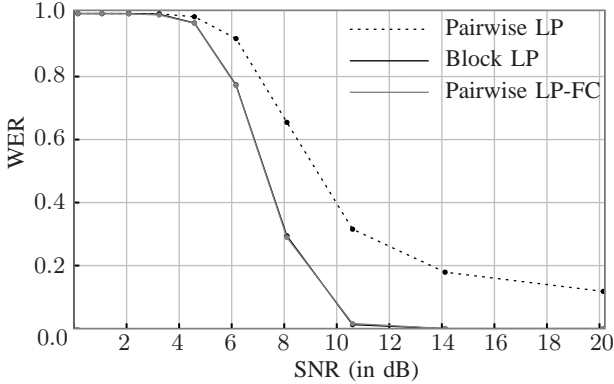} 
\caption{\label{fig:PLP_BRLP_PairwiseLPwithCycles_ISI=0.2} 
Figure shows simulation results for the performance of various LP detectors.
We plot  WER versus SNR for $h_{i,j}=0.2$ $\forall \, 1\leq i,j \leq N$ and
$\sigma=\{0.1,...,1.0\}$.  The dashed curve depicts the performance of the Pairwise LP.
The Pairwise LP performs quite poorly. Also, whenever Pairwise LP fails, it is because the LP
could not close the duality gap (LP gave fractional solution).  The solid curve corresponds to Block LP.  The
 Block LP performs much better, especially in the high SNR regime. For all
simulations the Block LP output was integral, implying that Block LP did MAP decoding
for this case. The curve in gray denotes the Pairwise LP-FC detector (see
Section~\ref{sec:LPwithcycles}). We observe that the Pairwise LP-FC curve sits right on
the top of the Block LP curve and also gave an integer solution every trial. 
} 
\end{figure}

Figure~\ref{fig:PLP_BRLP_PairwiseLPwithCycles_ISI=0.2} shows  WER versus SNR. 
The WER is quite high and the Pairwise LP performs poorly. We observe
that whenever the Pairwise LP fails, it is because the LP could not close the
duality gap. I.e., the output of the Pairwise LP is fractional (and hence we
declare an error). 



\begin{remark}
An important remark at this juncture is that when we solve the Pairwise LP (and
any other LPs which will follow), we always add a very small random perturbation
to the potential functions. This allows us to break ties when there are multiple
integer solutions.  
\end{remark}

\section{Improved LP Detectors} From the above experiments
it seems clear that Pairwise LP performs poorly. 
Most of the failure is because the LP outputs a
fractional solution. In the high SNR regime we expect that MAP decoder should
perform reasonably well. Hence, we now focus on improving the LP relaxation so
that, at least in the high SNR regime, we recover the transmitted word.
In other words, we aim to reduce the duality gap.   

\subsection{Block Linear Program}
\begin{figure}[htp] \centering
\input{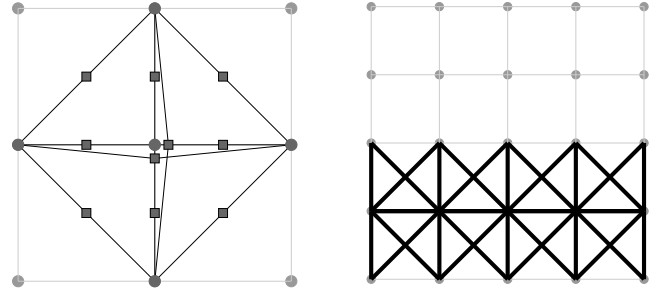} 
\vspace{-0.5cm}
\caption{\label{fig:block} 
 Figure on left shows a 5-clique. The figure on the right shows all the 5-cliques
with center on the second row. Boundary 5-cliques have the node
at the opposite end present as the fifth node. 
} 
\end{figure}

An immediate observation we make is that the pairwise interactions are not the
most natural cliques present in the factor graph. It is not hard to see that
the next-to-neighbor interactions (cf. Section~\ref{sec:IP}) introduces a 5-clique as shown in Figure~\ref{fig:block}. 
Thus the first enhancement, is to add all such 5-cliques
  to the LP. E.g., in a $9\times 9$ grid, there are 81 such 5-cliques which sit on each information bit. 
More precisely, we now have the following Block LP,
\begin{align}
&\min_{\bb} \sum_{i>j} \sum_{x_i, x_j} R_{ij} x_i x_j b_{ij}(x_i, x_j) - \sum_i\sum_{x_i}
{\sf h}_i x_i b_i(x_i) \nonumber \\
& \text{s.t.} \quad \forall i>j:\quad \sum_{x_i, x_j}b_{ij}(x_i, x_j)=1, \nonumber \\
& \forall i>j \quad \forall x_i, x_j:\quad
b_i(x_i)=\sum_{x_j}b_{ij}(x_i, x_j) \nonumber \\ 
& \hspace{3.0cm} b_j(x_j)=\sum_{x_i}b_{ij}(x_i,
x_j). \nonumber \\ 
& \forall C \quad \forall i,j \in C:\quad
b_{ij}(x_i,x_j)=\sum_{x_C\setminus x_i,x_j}b_{C}(x_C) \nonumber
\\ 
&0\leq b_i(x_i) \leq 1, \quad \forall i, \quad
 0\leq b_{ij}(x_i, x_j) \leq 1, \quad \forall i,j, \nonumber \\
& 0\leq b_{C}(x_C) \leq 1, \quad \forall C,\,\,C\text{ is a 5-clique}.
\nonumber
\end{align}

When we add all the 5-cliques to the LP, we have to make sure that they are
consistent (marginalization condition) across any intersections with other
cliques. It is not hard to see that any two 5-cliques intersect along an
edge of the clique. Thus the intersections are pairwise cliques.  Also, the
 5-cliques include the already present pairwise potentials as sub-cliques.
Hence, we have the above consistency condition between a 5-clique and all of
its constituent pairwise cliques. This relaxation resembles the GBP approach of \cite{ShWeShWe04}. 

\subsection{Experiments with Blockwise Linear Program} We consider the same
setup as in Section~\ref{sec:exptwithPLP} but with 
the Block LP.  Figure~\ref{fig:PLP_BRLP_PairwiseLPwithCycles_ISI=0.2} shows the WER
versus the SNR (in dB) for both the Pairwise LP and Block LP.  We see that the WER for the
 Block LP is much better than Pairwise LP for high SNR regime. In fact, in the high SNR
regime every simulation trial was correctly solved by the Block LP.  We also
observe that for every simulation (i.e., for every SNR), the LP outputs an
integral solution for each of the 2000 simulations. However, the output
information word is not always the transmitted word (e.g., when WER is
non-zero).  We conclude that, for this case, the Block LP is doing
MAP decoding.

\section{LP Detectors using Frustrated Subgraphs}\label{sec:LPwithFC}
Although the Block LP served our purpose of providing an optima low complexity 
detector, it seems adding all the 5-cliques is unnecessary. 
 In this section we investigate if there are ```smaller'' optimal LPs. 

Our approach is to adaptively add constraints to the LP which, simultaneously,
reduce the duality gap and are tractable (i.e., the number of such additional
constraints are small and also each constraint involves only a small number of
variables).  Such approaches, which try to get rid off the fractional solution
(or make the LP polytope tighter), have been used to improve the LP decoding of
LDPC codes \cite{TagSie08, DimWain09, Burshtein10}.  In \cite{DimWain09}, the
LP is enhanced by eliminating the facet containing the fractional solution. In
\cite{TagSie08, Burshtein10}, extra constraints are added by combining parity
checks which correspond to violated constraints to improve the LP performance.
Although our approach is in the same spirit, the main ideas have their origins
in \cite{JJthesis08} and \cite{JMW07}. Similar ideas have been independently used in \cite{SonJaa07, KoPaTz}. Before we describe the basic idea
let us first define the notion of a {\em frustrated graph}.

\begin{definition}[Frustrated Graph] Consider a constraint
satisfaction problem (CSP) defined on $n$ binary (boolean) variables, $\x$, and
$m$ check nodes. For each constraint node $\alpha$ there are
only certain configurations of $x_{\alpha}$ which satisfy it.  
Then, we say that the graph is {\em frustrated} if and only if there
is no assignment of $\x$ which satisfies all the constraint nodes
simultaneously. \qed 
\end{definition}

Let us now define a frustrated graph for our set-up. Assume that
the output of Pairwise LP is a fractional solution, i.e., we have a duality gap.
Consider all the potential functions (which have at least two variables) and
their LP beliefs. E.g., consider one of the 5-cliques, say $C$, and its beliefs
$b_C(x_C)$. We say that a configuration of $x_C$ satisfies $C$, if it has a
non-zero belief, i.e., $b_C(x_C)>0$. If the corresponding belief is zero, then
we say that it does not satisfy $C$. In other words, the set of configurations
which satisfy the potential function correspond to the support set of the
belief.  

\begin{lemma}
If there exists a frustrated subgraph, then there
 is a duality gap. 
\end{lemma}
\begin{proof}
Indeed, suppose on the contrary there was no duality gap.
This implies that the output of the LP is integer. I.e., all the beliefs (on
singleton potentials as well as higher order potentials) have only one
configuration with belief equal to 1 (rest being equal to zero). Consider any
subset of potential functions, $\mathcal{C}=\{C_1, C_2,\dots, C_r\}$. Let
$x^*_{C_i}$ denote the configuration such that $b_{C_i}(x^*_{C_i})=1.0$. We
claim that $\cup_i x^*_{C_i}$ satisfies the CSP represented by $\mathcal{C}$. 
 This follows from the consistency imposed by the LP (between any higher
order potential function and singleton potential functions). Thus no subgraph is frustrated.
\end{proof}

Now if we add a frustrated subgraph as a constraint in our LP, then we ensure
that this subgraph cannot be frustrated when we resolve the LP.  In
\cite{JJthesis08} it was found empirically that the random field ising model
could typically be solved (duality gap closed) by adding frustrated cycles
(cycles with odd number of frustrated interactions) arising in the LP solution.
It is also known from Barahona's work (see references within \cite{SonJaa07})
that adding cycles is sufficient to solve the zero-field planar ising model.

To ensure that the subgraph we add as a constraint to the LP becomes consistent
(or is not frustrated), we need to add all its maximal cliques and their
intersections to the LP. More precisely, we add the maximal cliques of the
junction tree\footnote{ See \cite{JJthesis08} for a discussion on Junction trees. It can
be shown that running LP on the junction tree of a graph is optimal (equal to
the IP).  The complexity of the LP grows exponentially in the size of the
maximal clique, which is the tree-width of the graph. Hence we focus on finding
frustrated subgraphs of small tree-width which keeps the LP tractable.} of that
subgraph as beliefs to the LP.

The main challenge that remains is to find a frustrated subgraph (with low
tree-width) in tractable time. In general, it is hard to find an arbitrary
subgraph which is frustrated. As a result, we focus on finding frustrated
cycles of the graph. This is a tractable problem and uses the implication graph
method (to solve 2SAT problem) of \cite{APT79, JJthesis08}. We describe it
briefly here, for details see Appendix B in \cite{JJthesis08}. Consider all the
two-projections of all the potential functions. I.e., for any $b_{C}(x_C)$
consider all the $b_{ij}(x_i, x_j) \,\,\forall\,\,i,j \in C$.  In the
implication graph each node $i$ is present as $i_+$ (for $x_i=0$) and $i_-$
(for $x_i=1$). There is a directed edge present between $i$ and $j$ which
represents the logical implication obtained from the potential
$b_{ij}(x_i,x_j)$. To generate this logical implication, consider the set $T$
of configurations of $(x_i, x_j)$ which render $b_{ij}(x_i,x_j)>0$ and can
introduce inconsistency.  Thus, $T$ is any of the following $(01,10),
(01,10,11), (01,10,00), (00,11), (00,11,10)$ and $(00,11,01)$.  Now one can
draw the directed edges using this $T$. E.g., suppose that LP outputs beliefs such that
$b_{ij}(0,1)>0, b_{ij}(1,0)>0, b_{ij}(1,1)>0, b_{i,j}(0,0)=0$
 then $T=(01,10,11)$ which would imply a directed edge
from $i_+\to j_-$ and $j_+ \to i_-$.  Then a frustrated cycle is defined to be
a directed cycle or path which visits both $i_+$ and $i_-$ for any $i$ and one
can find all such cycles and paths in linear time.  

\subsection{Experiments using Frustrated Cycles}\label{sec:LPwithcycles}

The set-up is exactly same as previous two experiments. The detector, which we call it Pairwise LP-FC, is as follows. \\

\begin{center}\framebox[0.95\columnwidth]{
\begin{minipage}{0.85\columnwidth}
\vspace{1mm}
\begin{enumerate}
\item  Run the Pairwise LP. Go to step 4).
\item  If the output is fractional, find all frustrated cycles (FC). For every FCs, add all the maximal cliques of its Junction tree to the LP. This ensures that we only add triangles. 
\item  Rerun the Pairwise LP.                                                    
\item If output is integral, stop else go to 2).                    
\end{enumerate}
\vspace{1mm}
\end{minipage}}
\end{center}
\vspace{3mm}
  We observe in Figure~\ref{fig:PLP_BRLP_PairwiseLPwithCycles_ISI=0.2}
that Pairwise LP-FC performs much better than the Pairwise LP and has the
same performance as the Block LP. Furthermore, Pairwise LP-FC gave an integer
output on every occasion. Thus in this case, Pairwise LP-FC 
 does MAP decoding. We also remark that the number of triangles added is
roughly 500 for each trial. This is much less than the total triangles
present in the graph ($= 85320$). Also, the step 2) above is run only once, if it is required.  

\subsection{Complexity of Block LP versus Pairwise LP with Cycles} We measure
the complexity of the LP by the number of nonzero entries in the LP constraint
matrix. In Table~\ref{tab:complexity}, the Pairwise LP-FC entries correspond to
the average (over 2000 simulations) number of nonzero entries in the matrix.
From the Table~\ref{tab:complexity} we see that, on an average, the Pairwise LP-FC has around
half the number of nonzero entries in the matrix when compared to the same in
Block LP.   Thus, the Pairwise
LP-FC is a ``smaller'' (more sparse), on average, when compared to the Block LP, with the same
performance. Also, the maximum nonzero entries (happens when step 2) is called) for Pairwise LP-FC is close to the Block LP one.  

\begin{table} \centering 
\begin{tabular}{|c|c|c|c|}
\hline
SNR & Block LP & Pairwise LP-FC (avg) & Pairwise LP-FC (max)\\
\hline
$20.6446$ & $4\times 10^4$ & $1.0966\times 10^4$ & $4.3578\times 10^4$\\ 
$14.6240$ & $4\times 10^4$ & $1.2163\times 10^4$ & $4.2258\times 10^4$\\
$11.1022$ & $4\times 10^4$ & $1.4887\times 10^4$ & $4.1202\times 10^4$\\
$8.6034 $ & $4\times 10^4$ & $1.8523\times 10^4$ & $4.1070\times 10^4$\\
$6.6652 $ & $4\times 10^4$ & $2.1440\times 10^4$ & $3.9618\times 10^4$\\
$5.0816 $ & $4\times 10^4$ & $2.2250\times 10^4$ & $4.2258\times 10^4$\\
$3.7426 $ & $4\times 10^4$ & $2.2162\times 10^4$ & $3.6758\times 10^4$\\
$2.5828 $ & $4\times 10^4$ & $2.1622\times 10^4$ & $3.9618\times 10^4$\\
$1.5597 $ & $4\times 10^4$ & $2.0143\times 10^4$ & $3.4558\times 10^4$\\
$0.6446 $ & $4\times 10^4$ & $1.9363\times 10^4$ & $3.5350\times 10^4$
\\ \hline 
\end{tabular} 
\caption{ \label{tab:complexity} Complexity comparison of Block LP and Pairwise LP-FC}
\vspace{-0.5cm}
\end{table}
\section{Discussion}
In this paper we develop channel detectors for the 2D ISI channel based
on LP. Although the Pairwise LP performs poorly,  both the Block LP and 
Pairwise LP-FC do MAP decoding.  As we mentioned
before, the advantage of LP detectors over GBP based detectors is
that the LP detectors provide a MAP certificate. Another advantage is that one
can formulate a systematic framework for improving the performance of LP
detectors. As mentioned in \cite{ShWeShWe04},
to date no systematic method of choosing regions (for the GBP algorithm) in a
general graph exists in order to improve the performance.
We end with possible open questions. \\
(i) In this work we consider only the low interference regime of $h_{i,j}=0.2$. It will be interesting to study the performance of the detectors when we vary the interference strength. \\ 
(ii) An interesting research
direction is to combine coding (or precoding) with channel detection and to
develop a joint decoder and detector based on LP. \\
(iii) Another open question is to study LP detectors when there is non-linear ISI \cite{SiSu05}. 
  This would introduce higher order interactions in the factor graph. \\
(iv) LP is also used to decode LDPC codes when transmitting over binary-input memoryless channels \cite{RiU08}. An interesting question is to see if the LP decoder \cite{FWK05} enhanced using frustrated
cycles/subgraphs can lead to improvement in decoding thresholds. 

\section{Acknowledgments} Our work at LANL was carried out under the auspices
of the National Nuclear Security Administration of the U.S. Department of
Energy at Los Alamos National Laboratory under Contract No.  DE-AC52-06NA25396.
SK acknowledges support of NMC via the NSF collaborative grant CCF-0829945 on
``Harnessing Statistical Physics for Computing and Communications.'' 

\bibliographystyle{IEEEtran} 
\bibliography{lanl}

\newcommand{\SortNoop}[1]{}
\begin{thebibliography}{10}
\providecommand{\url}[1]{#1}
\csname url@rmstyle\endcsname
\providecommand{\newblock}{\relax}
\providecommand{\bibinfo}[2]{#2}
\providecommand\BIBentrySTDinterwordspacing{\spaceskip=0pt\relax}
\providecommand\BIBentryALTinterwordstretchfactor{4}
\providecommand\BIBentryALTinterwordspacing{\spaceskip=\fontdimen2\font plus
\BIBentryALTinterwordstretchfactor\fontdimen3\font minus
  \fontdimen4\font\relax}
\providecommand\BIBforeignlanguage[2]{{%
\expandafter\ifx\csname l@#1\endcsname\relax
\typeout{** WARNING: IEEEtran.bst: No hyphenation pattern has been}%
\typeout{** loaded for the language `#1'. Using the pattern for}%
\typeout{** the default language instead.}%
\else
\language=\csname l@#1\endcsname
\fi
#2}}

\bibitem{Co03}
W.~Coene, ``Two-dimensional optical storage,'' in \emph{Tech. Dig. Opt. Data
  Storage (ODS) Conf.}, Vancouver, Canada, 2003, pp. 90--92.

\bibitem{ImmCo03}
A.~H.~J. Immink, W.~M.~J. Coene, van~der Lee A.~M., C.~Busch, A.~P. Hekstra,
  J.~W.~M. Bergmans, J.~Riani, S.~J. L.~V. Beneden, and T.~Conway, ``Signal
  processing and coding for two-dimensional optical storage,'' in \emph{Proc.
  of GLOBECOM}, vol.~7, San Francisco, USA, Dec. 2003, pp. 3904--3908.

\bibitem{Sie06}
P.~H. Siegel, ``Information-theoretic limits of two-dimensional optical
  recording channels,'' in \emph{Optical Data Storage, Proceedings of SPIE},
  Montreal, Canada, Apr. 2006.

\bibitem{Kur08}
B.~M. Kurkoski, ``Towards efficient detection of two-dimensional intersymbol
  interference channels,'' \emph{IEICE Transactions on Fundamentals of
  Electronics, Communications and Computer Sciences}, vol. E91, no.~10, Oct.
  2008.

\bibitem{For72}
G.~D. Forney, Jr., ``Maximum-likelihood sequence estimation of digital
  sequences in the presence of intersymbol interference,'' \emph{IEEE Trans.
  Inform. Theory}, vol.~18, no.~3, pp. 363--378, May 1972.

\bibitem{OR06}
E.~Ordentlich and R.~M. Roth, ``On the computation complexity of 2d
  maximum-likelihood sequence detection,'' July 2006, technical Report
  HPL2006-69.

\bibitem{MaWo03a}
M.~Marrow and J.~Wolf, ``Iterative detection of 2-dimensional isi channels,''
  in \emph{Proc. of the IEEE Inform. Theory Workshop}, Apr. 2003.

\bibitem{MaWo03b}
------, ``Detection of 2-dimensional signals in the presence of isi and
  noise,'' in \emph{Proc. International Symposium on Information Theory and its
  Applications}, Oct. 2003, pp. 891--894.

\bibitem{ChAnCh01}
K.~M. Chugg, A.~Anastasopoulos, and X.~Chen, ``Iterative detection: Adaptivity,
  complexity reduction, and applications,'' \emph{Kluwer {A}cademic
  {P}ublishers}, 2001.

\bibitem{Kris98}
R.~Krishnamoorthi, ``Two-dimensional viterbi-like algorithms,'' 1998, master's
  Thesis.

\bibitem{Weeks00}
W.~Weeks, ``Full-surface data storage,'' 2000, phD Thesis.

\bibitem{HekCoImm07}
A.~Hekstra, W.~Coene, and A.~Immink, ``Refinements of multi-track viterbi,''
  \emph{IEEE Trans. Magn.}, vol.~43, no.~7, pp. 3333--3339, 2007.

\bibitem{WuSuSiIn03}
Y.~Wu, J.~A. O’Sullivan, N.~Singla, and R.~Indeck, ``Iterative detection and
  decoding for separable two-dimensional intersymbol interference,'' \emph{IEEE
  Trans. Magn.}, vol.~39, no.~4, pp. 2115--2120, July 2003.

\bibitem{SiSuInWu02}
N.~Singla, J.~O’Sullivana, R.~Indeck, and Y.~Wu, ``Iterative decoding and
  equalization for 2-d recording channels,'' \emph{IEEE Trans. Magn.}, vol.~38,
  no.~5, pp. 2328--2330, Sept. 2002.

\bibitem{SiSu05}
N.~Singla and J.~O’Sullivan, ``Joint equalization and decoding for nonlinear
  two-dimensional intersymbol interference channels,'' in \emph{Proc. of the
  IEEE Int. Symposium on Inform. Theory}, Adelaide, Australia, Sept. 2005.

\bibitem{ShShSha05}
O.~Shental, N.~Shental, and S.~Shamai, ``On the achievable information rates of
  ﬁnite-state input two-dimensional channels with memory,'' in \emph{Proc. of
  the IEEE Int. Symposium on Inform. Theory}, Adelaide, Australia, Sept. 2005,
  pp. 2354--2358.

\bibitem{ShWeShWe04}
O.~Shental, A.~Weiss, N.~Shental, and Y.~Weiss, ``Generalized belief
  propagation receiver for near-optimal detection of two-dimensional channels
  with memory,'' in \emph{Proc. of the IEEE Inform. Theory Workshop}, San
  Antonio, USA, Oct. 2004, pp. 225--229.

\bibitem{FWK05}
J.~Feldman, M.~J. Wainwright, and D.~R. Karger, ``Using linear programming to
  decode binary linear codes,'' \emph{IEEE Trans. Inform. Theory}, vol.~51,
  no.~3, Mar. 2005.

\bibitem{TagSie08}
M.-H. Taghavi and P.~Siegel, ``Adaptive methods for linear programming
  decoding,'' \emph{IEEE Trans. Inform. Theory}, vol.~54, no.~12, pp.
  5396--5410, 2006.

\bibitem{DimWain09}
A.~Dimakis, A.~Gohari, and M.~Wainwright, ``Guessing facets: Polytope structure
  and improved lp decoder,'' \emph{IEEE Trans. Inform. Theory}, vol.~55, no.~8,
  pp. 3479--3487, 2009.

\bibitem{Burshtein10}
D.~Burshtein and I.~Goldenberg, ``Improved linear programming decoding and
  bounds on the minimum distance of {LDPC} codes,'' in \emph{Proc. of the IEEE
  Inform. Theory Workshop}, Dublin, Ireland, Aug. 2010.

\bibitem{JJthesis08}
J.~Johnson, ``Convex relaxation methods for graphical models: Lagrangian and
  maximum entropy approaches,'' 2008, {P}h{D} Thesis.

\bibitem{JMW07}
J.~Johnson, D.~Malioutov, and A.~Willsky, ``Lagrangian relaxation for map
  estimation in graphical models,'' in \emph{Proc. of the Allerton Conf. on
  Commun., Control, and Computing}, Sept. 2007.

\bibitem{SonJaa07}
D.~Sontag and T.~Jaakkola, ``New outer bounds on the marginal polytope,'' in
  \emph{Neural Information Processing Systems (NIPS)}, Dec. 2007.

\bibitem{KoPaTz}
N.~Komodakis, N.Paragios, and G.~Tziritas, ``{MRF} energy minimization and
  beyond via dual decomposition,'' \emph{IEEE Trans. on Pattern Analysis and
  Machine Intelligence}, (in press).

\bibitem{APT79}
B.~Aspvall, M.~Plass, and R.~Tarjan, ``A linear-time algorithm for testing the
  truth of certain quantified boolean formulas,'' in \emph{Information
  Processing Letters}, 1979, 8(3).

\bibitem{RiU08}
T.~Richardson and R.~Urbanke, \emph{Modern Coding Theory}.\hskip 1em plus 0.5em
  minus 0.4em\relax Cambridge University Press, 2008.

\end{thebibliography}
\end{document}